\documentclass[a4paper,11pt]{article}

\usepackage{fancyhdr}

\usepackage[T1]{fontenc} 
\usepackage[utf8]{inputenc} 
\usepackage{lipsum} 

\usepackage[top=3cm, bottom=3cm, left=3.8cm, right=3.8cm, heightrounded,
            marginparwidth=2.5cm, marginparsep=0.7cm]{geometry} 

\usepackage{changepage} 

\usepackage[shortlabels]{enumitem} 

\usepackage[square,numbers,merge,comma,sort&compress]{natbib} 
\makeatletter
\def\NAT@spacechar{\,}  
\makeatother

\usepackage{amsmath,amssymb,amsfonts,amsthm}
\usepackage{mathtools} 
\usepackage{slashed,cancel}
\usepackage{comment}   
\usepackage{lmodern}   
\usepackage{relsize}
\usepackage{setspace}  
\usepackage{moresize}  
\usepackage{epsfig}
\usepackage{latexsym}
\usepackage{mathrsfs,calligra,aurical} 
\usepackage{calc}
\usepackage{float}
\usepackage{appendix}
\usepackage{xargs}
\usepackage{extarrows}
\usepackage{empheq}

\usepackage{soul}

\usepackage[blocks]{authblk}  

\usepackage[fulladjust]{marginnote}%

\usepackage{graphicx} 
\graphicspath{{Figures/}} 

\usepackage[labelsep=colon]{caption}  
\usepackage[labelsep=colon,aboveskip=10pt,belowskip=10pt]{subcaption}
\usepackage{array,multirow,makecell,booktabs}  
\newcolumntype{M}[2]{>{\centering\arraybackslash$}#1{#2\linewidth}<{$}}
\newcolumntype{T}[2]{>{\centering\arraybackslash}#1{#2\linewidth}<}
\newcolumntype{R}{>{\arraybackslash$}r<{$}}
\newcolumntype{L}{>{\arraybackslash$}l<{$}}
\def\thinrule{\midrule[0.0001pt]}

\usepackage{titlesec}

\titlespacing*{\section}{0pt}
              {7.5ex plus 1ex minus .2ex}{2.ex plus .2ex minus .2ex}
\titlespacing*{\subsection}{0pt}
              {3.5ex plus 1ex minus .2ex}{1.25ex plus .2ex minus .2ex}
\titlespacing*{\subsubsection}{0pt}
              {3.25ex plus 1ex minus .2ex}{1.25ex plus .2ex minus .2ex}

\titleclass{\subsubsubsection}{straight}[\subsubsection]
\newcounter{subsubsubsection}[subsubsection]
\renewcommand\thesubsubsubsection{\thesubsubsection.\arabic{subsubsubsection}}
\titleformat{\subsubsubsection}{\normalfont\large}{\thesubsubsubsection}{1em}{}
\titlespacing*{\subsubsubsection}
              {0pt}{3ex plus 1ex minus .2ex}{1ex plus .2ex minus .2ex}

\makeatletter
\renewcommand\paragraph{\@startsection{paragraph}{5}{\z@}%
      {2.5ex \@plus1ex \@minus.2ex}{-1em}{\bfseries}}
\renewcommand\subparagraph{\@startsection{subparagraph}{6}{\z@}
      {2.5ex \@plus1ex \@minus .2ex}{-1em}{\bfseries}}
\def\toclevel@subsubsubsection{4} 
\def\toclevel@paragraph{5}
\def\toclevel@subparagraph{6}
\@addtoreset{subsubsubsection}{section} 
\@addtoreset{subsubsubsection}{subsection}
\@addtoreset{paragraph}{subsubsubsection}
\makeatother

\usepackage{titletoc}

\addtocontents{toc}{\addvspace{-0.75em}}  

\titlecontents{section}
  [1.25em] {\addvspace{0.7em plus 0pt}\small}
  {\thecontentslabel\hspace{0.75em}}{}
  {\hspace{0.5em}\titlerule*[0.5em]{.}\contentspage}
  [\addvspace{0.0em plus 0pt}]

\titlecontents{subsection}
  [2.75em] {\addvspace{0.075em plus0pt}\fns}
  {\thecontentslabel\hspace{0.75em}}{\thecontentslabel\hspace{0.75em}}
  {\hspace{0.5em}\titlerule*[0.5em]{.}\small\contentspage}
  [\addvspace{0.075em plus 0pt}]

\usepackage{environ}
\makeatletter
\NewEnviron{subalign}[1]{
\begin{subequations}\label{#1}
\begin{align} \BODY \end{align}
\end{subequations}      }
\makeatother
\makeatletter
\newenvironment{subeqs}%
{\begingroup%
\setlength{\abovedisplayskip}{10pt plus 4pt minus 9pt}%
\setlength{\abovedisplayshortskip}{0pt plus 2pt minus 2pt}%
\setlength{\belowdisplayskip}{12pt plus 3pt minus 9pt}%
\setlength{\belowdisplayshortskip}{7pt plus 3pt minus 4pt}%
\begin{subequations}%
}%
{\end{subequations}\ignorespacesafterend
\endgroup}%
\makeatother

\makeatletter
\newenvironment{subeqsds}[2]%
{\begingroup%
\setlength{\abovedisplayskip}{{#1}pt plus 2pt minus 9pt}%
\setlength{\abovedisplayshortskip}{0pt plus 0pt minus 2pt}%
\setlength{\belowdisplayskip}{{#2}pt plus 3pt minus 9pt}%
\setlength{\belowdisplayshortskip}{7pt plus 3pt minus 4pt}%
\begin{subequations}%
}%
{\end{subequations}\ignorespacesafterend
\endgroup}%
\makeatother

\makeatletter
\newenvironment{equationds}[2]%
{\begingroup%
\setlength{\abovedisplayskip}{{#1}pt plus 3pt minus 9pt}%
\setlength{\abovedisplayshortskip}{0pt plus 3pt}%
\setlength{\belowdisplayskip}{{#2}pt plus 3pt minus 9pt}%
\setlength{\belowdisplayshortskip}{7pt plus 3pt minus 4pt}%
\begin{equation}%
}%
{\end{equation}\ignorespacesafterend
\endgroup}%
\makeatother

\usepackage{xcolor}
\definecolor{Green}{rgb}{0.05, 0.45, 0.25}
\definecolor{dogwoodrose}{rgb}{0.8, 0.1, 0.55}
\definecolor{RRed}{rgb}{0.65, 0.1, 0.5}

\usepackage{bm}  
\usepackage{dsfont}  

\DeclareMathAlphabet{\mathpzc}{OT1}{pzc}{m}{it}

\DeclareMathAlphabet{\mathcal}{OMS}{cmsy}{m}{n}
\DeclareSymbolFontAlphabet{\Scr}{rsfs}

\DeclareMathAlphabet{\mathbold}{U}{BOONDOX-ds}{m}{n}
\SetMathAlphabet{\mathbold}{bold}{U}{BOONDOX-ds}{b}{n}

\DeclareMathAlphabet{\mathcalboondox}{U}{BOONDOX-calo}{m}{n}
\SetMathAlphabet{\mathcalboondox}{bold}{U}{BOONDOX-calo}{b}{n}
\DeclareMathAlphabet{\mathbcalboondox}{U}{BOONDOX-calo}{b}{n}

\def\linkcol{RRed}



\usepackage[breaklinks=true,backref=page]{hyperref}
\hypersetup{
    pdfpagemode={UseNone},
    pdfstartview={FitH},
    colorlinks=true,
    bookmarks=true,
    bookmarksnumbered=true,
    plainpages,
    a4paper,
    linktoc=page,
    citecolor=blue,
    filecolor=black,
    linkcolor=\linkcol,
    urlcolor=Green,
}
\renewcommand*{\backref}[1]{}
\renewcommand*{\backrefalt}[4]{%
\ifcase #1 %
\relax
\or
~{\small [\textsc{p.~\fns{\!#2}}]}
\else
~{\small [\textsc{p.~\fns{\!#2}}]}%
\fi}

\newif\ifbackrefshowonlyfirst
\backrefshowonlyfirsttrue
\makeatletter
\let\BR@direct@old@hyper@natlinkstart\hyper@natlinkstart
\renewcommand*{\hyper@natlinkstart}{\phantomsection\BR@direct@old@hyper@natlinkstart}%
\let\BR@direct@oldBR@citex\BR@citex
\renewcommand*{\BR@citex}{\phantomsection\BR@direct@oldBR@citex}%
\long\def\hyper@page@BR@direct@ref#1#2#3{\hyperlink{#3}{#1}}
\ifx\backrefxxx\hyper@page@backref
    \let\backrefxxx\hyper@page@BR@direct@ref
    \ifbackrefshowonlyfirst
        
    \fi
\else
    \ifbackrefshowonlyfirst
    \fi
\fi
\RequirePackage{etoolbox}
\patchcmd{\Hy@backout}{Doc-Start}{\@currentHref}{}{\errmessage{I can't seem to patch backref}}
\makeatother

\usepackage{footnotebackref}
\usepackage{hypernat} 

\def\+{~+~}
\def\-{~-~}
\def\={~=~}

\newcommand\fns{\footnotesize}

\newcommand\qRq{\quad\Rightarrow\quad}

\newcommand\dd{\partial}

\newcommand\epsz{\varepsilon_\ms{0}}
\newcommand\epsg{\varepsilon_\textrm{g}}
\newcommand\mug{\mu_\textrm{g}}
\newcommand\muz{\mu_\ms{0}}

\newcommand\e{\textrm{e}}

\newcommand\DD{\mathcal{D}}
\newcommand\Tct{$T_\textrm{c}$~}
\newcommand\Tc{T_\textrm{c}}
\newcommand\Hc{\textrm{H}_\textrm{c}}
\newcommand\E{\mathbf{E}}
\newcommand\B{\mathbf{B}}
\newcommand\A{\mathbf{A}}
\newcommand\HH{\mathbf{H}}
\newcommand\Eg{\mathbf{E}_\textrm{g}}
\newcommand\Bg{\mathbf{B}_\textrm{g}}
\newcommand\Ag{\mathbf{A}_\textrm{g}}
\newcommand\Ee{\mathbf{E}_\textrm{e}}
\newcommand\Be{\mathbf{B}_\textrm{e}}
\newcommand\Ae{\mathbf{A}_\textrm{e}}
\newcommand\jj{\mathbf{j}}
\newcommand\jg{\mathbf{j}_\textrm{g}}

\newcommand\ns{n_\textrm{s}}
\newcommand\vs{v_\textrm{s}}
\newcommand\lambdag{\lambda_\textrm{g}}
\newcommand\lambdae{\lambda_\textrm{e}}
\newcommand\rhog{\rho_\textrm{g}}
\newcommand\phig{\phi_\textrm{g}}
\newcommand\n{\mathbf{n}}
\newcommand\gstar{g_\star}
\newcommand\PP{\mathcal{P}(x)}
\newcommand\vF{v_\textsc{f}}
\newcommand\Cnq{\mathcal{C}^{2}_n}
\newcommand\Qn{\mathcal{Q}_n}
\newcommand\wn{\omega_n}
\newcommand\Rn{\mathcal{R}_n}
\newcommand\Sn{\mathcal{S}_n}
\newcommand\mt{\mathrm{m}}
\newcommand\s{\mathrm{s}}
\newcommand\K{\mathrm{K}}
\newcommand\GN{\mathrm{G}_\ms{\textsc{n}}}

\newcommand\LL{\mathscr{L}}

\providecommand{\abs}[1]{\lvert#1\rvert}

\newcommand{\ms}{\mathsmaller}

\newcommandx{\tts}[1]{\text{\textsmaller{#1}}}
\newcommandx{\dm}[1][1=\mu,usedefault]{\partial_{#1}}
\newcommandx{\dmup}[1][1=\mu,usedefault]{\partial^{#1}}
\newcommandx{\subm}[2][1=p,2=A,usedefault]{{#1}_{\!\mathsmaller{#2}}}
\newcommandx{\subt}[2][1=p,2=A,usedefault]{{#1}_\text{\textsmaller{#2}}}
\newcommandx{\supm}[2][1=p,2=A,usedefault]{{#1}^{\!\mathsmaller{#2}}}
\newcommandx{\supt}[2][1=p,2=A,usedefault]{{#1}^\text{\textsmaller{#2}}}
\newcommandx{\subpt}[3][1=p,2=A,3=B,usedefault]{{#1}^\text{\textsmaller{#3}}_\text{\textsmaller{#2}}}
\newcommandx{\subpm}[3][1=p,2=A,3=B,usedefault]{{#1}^{\mathsmaller{#3}}_{\mathsmaller{#2}}}
\newcommandx{\gmetr}[2][1=\mu,2=\nu,usedefault]{g_{{#1}{#2}}}
\newcommandx{\invgmetr}[2][1=\mu,2=\nu,usedefault]{g^{{#1}{#2}}}
\newcommandx{\emetr}[2][1=\mu,2=\nu,usedefault]{\eta_{{#1}{#2}}}
\newcommandx{\invemetr}[2][1=\mu,2=\nu,usedefault]{\eta^{{#1}{#2}}}
\newcommandx{\hmetr}[2][1=\mu,2=\nu,usedefault]{h_{{#1}{#2}}}
\newcommandx{\invhmetr}[2][1=\mu,2=\nu,usedefault]{h^{{#1}{#2}}}
\newcommandx{\bhmetr}[2][1=\mu,2=\nu,usedefault]{\bar{h}_{{#1}{#2}}}
\newcommandx{\binvhmetr}[2][1=\mu,2=\nu,usedefault]{\bar{h}^{{#1}{#2}}}
\newcommandx{\hud}[2][1=\mu,2=\nu,usedefault]{{h^{#1}}_{\!\!#2}}
\newcommandx{\Ruddd}[4][1=\sigma,2=\mu,3=\lambda,4=\nu,usedefault]{{R^{#1}}_{\!{#2}{#3}{#4}}}
\newcommandx{\Gam}[3][1=\lambda,2=\mu,3=\nu,usedefault]{{\Gamma^{#1}}_{\!{#2}{#3}}}
\newcommandx{\Gamd}[3][1=\mu,2=\nu,3=\lambda,usedefault]{\Gamma_{{#1}{#2}{#3}}}
\newcommandx{\Ricci}[2][1=\mu,2=\nu,usedefault]{R_{{#1}{#2}}}
\newcommandx{\GEinst}[2][1=\mu,2=\nu,usedefault]{G^{{}^\tts{(E)}}_{{#1}{#2}}}
\newcommandx{\Gscr}[3][1=\mu,2=\nu,3=\rho,usedefault]{\mathscr{G}_{{#1}{#2}{#3}}}
\newcommandx{\sh}[1][1=\alpha,usedefault]{\sinh\left(#1\right)}
\newcommandx{\ch}[1][1=\alpha,usedefault]{\cosh\left(#1\right)}
\newcommandx{\sech}[1][1=\alpha,usedefault]{\mathrm{sech}\left(#1\right)}
\newcommandx{\cosech}[1][1=\alpha,usedefault]{\mathrm{cosech}\left(#1\right)}

\title{%
       \vspace{-1.5cm}
       \begin{adjustwidth}{-0.5in}{-0.5in}
       \centering\boldmath\LARGE\bfseries%
       Superconductor in a weak static gravitational field
       \end{adjustwidth}%
       \bigskip
       }

\author{Giovanni Alberto Ummarino}
\affil{Politecnico di Torino, Dipartimento DISAT
      }
\affil{National Research Nuclear University MEPhI - Moscow Engineering
       Physics Institute
       }
\affil{\href{mailto:giovanni.ummarino@polito.it}{\texttt{giovanni.ummarino@polito.it}}
       }
\author{Antonio Gallerati}
\affil{Politecnico di Torino, Dipartimento DISAT
      }
\affil{Istituto Nazionale di Fisica Nucleare (INFN),
       Sezione di Torino, Italy
       }
\affil{\href{mailto:antonio.gallerati@polito.it}{\texttt{antonio.gallerati@polito.it}}
      }

\date{}

\begin{document}

\maketitle


\begin{abstract}
\noindent
We provide the detailed calculation of a general form for Maxwell and London equations that takes into account gravitational corrections in linear approximation. We determine the possible alteration of a static gravitational field in a superconductor making use of the time-dependent Ginzburg-Landau equations, providing also an analytic solution in the weak field condition. Finally, we compare the behavior of a \mbox{high-\Tct} superconductor 
with a classical low-\Tct superconductor, analyzing the values of the parameters that can enhance the reduction of the gravitational field.
\end{abstract}

\medskip

\tableofcontents



\pagebreak


\section{Introduction} \label{sec:Intro}
There is no doubt that the interplay between the theory of the gravitational field and superconductivity is a very intriguing field of
research, whose theoretical study has been involving many researchers for a long time \cite{DeWitt:1966yi,papini1967detection,Felch:1985pre,anandan1994relgra,anandan1977nuovo,anandan1984relthe,ross1983london,hirakawa1975super,rystephanick1973london,peng1991interaction,ciubotariu1996absence,agop1996gravitational,dinariev1987relativistic,minasyan1976londons,rothen1968application,li1991effects,peng1991electrodynamics,li1992gravitational,torr1993gravitoelectric}.
Podkletnov and Nieminem  declared the achievement of experimental evidence for a gravitational shielding in a high-\Tct superconductor (HTSC) \cite{podkletnov1992possibility,podkletnov1997weak}. After their announcement, other groups tried to repeat the experiment obtaining controversial results \cite{li1997static,de1995alternative,unnikrishnan1996does}, so that the question is still open.\par
In 1996, G.\ Modanese interpreted the results by Podkletnov and Nieminem in the frame of the quantum theory of General Relativity \cite{modanese1996theoretical,modanese1996role} but the complexity of the formalism makes it very difficult to extract quantitative
predictions.
Afterwards, Agop et al.\ wrote generalized Maxwell equations that simultaneously treat weak gravitational and electromagnetic fields \cite{agop2000local,agop2000some}.

\paragraph{Superfluid coupled to gravity.}
It is well known that, in general, the gravitational force is not influenced by any dielectric-type effect involving the medium. In the classical case, this is due to the absence of a relevant number of charges having opposite sign which, redistributing inside the medium, might counteract the applied field. On the other side, if we regard the medium as a quantum system, the probability of a graviton excitation of a medium particle is suppressed, due to the smallness of gravitational coupling. This means that any kind of shielding due to the presence of the medium can only be the result of an interaction with a different state of matter, like a Bose condensate or a more general superfluid.\par
The nature of the involved field is also relevant for the physical process. If the gravitational field itself is considered as classical, it is readily realized that no experimental device -- like the massive superconducting disk of the Podkletnov experiment \cite{podkletnov1992possibility,podkletnov1997weak} -- can influence the local geometry so much as to modify the measured sample weight. This means that the hypothetical shielding effect should consist of some kind of modification (or ``absorption'') of the field in the superconducting disk.\par
Since the classical picture is excluded, we need a quantum field description for the gravitational interaction \cite{modanese1996theoretical,modanese1996role}.
In perturbation theory the metric $\gmetr(x)$ is expanded in the standard way \cite{Wald:1984rg}
\begin{equation} \label{eq:metricweak}
\gmetr(x)\=\emetr+\hmetr(x)
\end{equation}
as the sum of the flat background $\emetr$ plus small fluctuations encoded in the $h_{\mu\nu}(x)$ component. The Cooper pairs inside the superconducting sample compose the Bose condensate, described by a bosonic field $\phi$ with non-vanishing vacuum expectation value $\phi_0=\langle0|\phi|0\rangle$.\par\smallskip
The Einstein-Hilbert Lagrangian has the standard form%
\footnote{we work in the ``mostly plus'' framework, $\eta=\mathrm{diag}(-1,+1,+1,+1)$, and set $c=\hbar=1$}%
\begin{equation}
\LL_\textsc{eh} \= \frac{1}{8\pi\GN}\,\left(R-2\,\Lambda\right)\;,
\end{equation}
where $R$ is the Ricci scalar and $\Lambda$ is the cosmological constant. The part of the Lagrangian describing the bosonic field $\phi$ coupled to gravity has the form:
\begin{equation}
\LL_\ms{\!\phi} \=
-\frac12\,\invgmetr\,\dm{\phi}^\ast\,\dm[\nu]\phi + \frac12\,m^2\,\phi^{\ast}\phi
\end{equation}
where $m$ is the mass of the Cooper pair \cite{modanese1996theoretical}.\par
If we expand the bosonic field as $\phi=\phi_0+\bar{\phi}$, one can consider the v.e.v.\ $\phi_0$ as an external source, related to the structure of the sample and external electromagnetic fields, while the $\bar{\phi}$ component can be included in the integration variables. The terms including the $\bar\phi$ components are related to graviton emission-absorption processes (which we know to be irrelevant) and can safely be neglected in $\LL_\ms{\!\phi}$\,. Perturbatively, the interaction processes involving the metric fluctuations and the condensate are of the form
\begin{equation}
\LL_\tts{int} ~\propto~
h^{\mu\nu}\,\dm{\phi_0}^\ast\,\dm[\nu]\phi_0
\label{eq:Lint}
\end{equation}
and give rise to (gravitational) propagator corrections, which are again irrelevant.\par
The total Lagrangian \;$\LL=\LL_\textsc{eh}+\LL_\ms{\!\phi}$\; contains a further coupling between $\gmetr$ and $\phi_0$, which turns out to be a contribution to the so-called intrinsic cosmological term given by $\Lambda$. Explicitly, the total Lagrangian can in fact be rewritten as
\begin{equation}
\LL\=\LL_\textsc{eh}+\LL_\ms{\!\phi}\=
\frac{1}{8\pi\GN}\,\left(R-2\,\Lambda\right)+\LL_\tts{int}
+\LL_0+\bar{\LL}_\ms{\bar{\phi}}\;,
\end{equation}
where $\bar{\LL}_\ms{\bar{\phi}}$ are the negligible contributions having at least one field $\bar{\phi}$ and where
\begin{equation}
\LL_0\=
-\frac12\,\dm{\phi_0}^\ast\,\dmup[\mu]\phi_0 + \frac12\,m^2\left|\phi_0\right|^2\;,
\end{equation}
that is, a Bose condensate contribution to the total effective cosmological term. This may produce slightly localized ``instabilities'' and thus an observable effect, in spite of the smallness of the gravitational coupling \eqref{eq:Lint}.\par
The above instabilities can be found in the superconductor regions where the condensate density is larger: in these regions, the gravitational field would tend to assume fixed values due to some physical cutoff, that prevents arbitrary growth. The mechanism is similar to classical electrostatics in perfect conductors, where the electric field is constrained to be globally zero within the sample. In the latter case, the physical constraint's origin is different (and is due to a charge redistribution), but in both cases the effect on field propagation and on static potential turns out to be a kind of partial shielding.\par
In accordance with the framework previously exposed, the superfluid density $\phi_0(x)$ is determined not only by the internal microscopic structure of the sample, but also by the same magnetic fields responsible for the Meissner effect and the currents in the superconductor. The high-frequency components of the magnetic field can also provide energy for the above gravitational field modification \cite{modanese1996theoretical}.\par
The previous calculation shows how Modanese was able to demonstrate, in principle, how a superfluid can determine a gravitational shielding effect.
In Sect.\ \ref{sec: Isotropic SC} we will quantify this effect by following a different approach, as the Ginzburg-Landau theory for a superfluid in an external gravitational field.

\section{Weak field approximation} \label{sec: Weak field}
\sloppy
Now we consider a nearly flat spacetime configuration, i.e.\ an approximation where the gravitational field is weak and where we shall assume eq.\ \eqref{eq:metricweak}, that is, the metric $\gmetr$ can be expanded as:
\begin{equation} \label{eq:gmetr}
\gmetr~\simeq~\emetr+\hmetr\;,
\end{equation}
where the symmetric tensor $\hmetr$ is a small perturbation of the flat Minkowski metric in the mostly plus convention%
\footnote{see Appendix~\ref{app:signconv} for definitions and sign conventions}%
, {$\emetr=\mathrm{diag}(-1,+1,+1,+1)$}. The inverse metric in the linear approximation is given by
\begin{equation}
\invgmetr~\simeq~\invemetr-\invhmetr\;.
\end{equation}
%

\subsection{Generalizing Maxwell equations}
If we consider an inertial coordinate system, to linear order in $\hmetr$ the connection is written as
\begin{equation}\label{eq:Gam}
\Gam[\lambda][\mu][\nu]~\simeq~\frac12\,\invemetr[\lambda][\rho]\,
     \left(\dm[\mu]\hmetr[\nu][\rho]+\dm[\nu]\hmetr[\rho][\mu]-\dm[\rho]\hmetr[\mu][\nu]\right)\;,
\end{equation}
The Ricci tensor (Appendix~\ref{app:signconv}) is given by the contraction of the Riemann tensor
\begingroup%
\setlength{\belowdisplayshortskip}{5pt plus 1pt}%
\setlength{\belowdisplayskip}{7pt plus 1pt minus 4pt}
\begin{equation}
\Ricci\=\Ruddd[\sigma][\mu][\sigma][\nu]\,.
\end{equation}
\endgroup
and, to linear order in $\hmetr$, it reads
\begin{equation} \label{eq:Ricci}
\begin{split}
\Ricci&~\simeq~\dm[\lambda]\Gam[\lambda][\mu][\nu]\+\dm[\mu]\Gam[\lambda][\lambda][\nu]\+\cancel{\Gamma\,\Gamma}\-\cancel{\Gamma\,\Gamma}\=\\
   &\=\frac12\,\left(\dm\dmup[\rho]\hmetr[\nu][\rho]+\dm[\nu]\dmup[\rho]\hmetr[\mu][\rho]\right)
      -\frac12\,\dm[\rho]\dmup[\rho]\hmetr-\frac12\,\dm\dm[\nu]h\=\\
   &\=\dmup[\rho]\dm[{(}\mu]\hmetr[\nu{)}][\rho]-\frac12\,\dd^2\hmetr-\frac12\,\dm\dm[\nu]h\;,
\end{split}
\end{equation}
where we have used eq.\ \eqref{eq:Gam} and where $h=\hud[\sigma][\sigma]$\,.\par\medskip
The Einstein equations have the form \cite{Wald:1984rg,misner1973gravitation}:
\begin{equation}
\GEinst\=\Ricci-\dfrac12\,\gmetr\,R
\=8\pi\GN\;T_{\mu\nu}\;,
\end{equation}
and the term with the Ricci scalar $R=\invgmetr\Ricci$ can be rewritten, in first-order approximation and using eq.\ \eqref{eq:Ricci}, as
\begin{equation}
\frac12\,\gmetr\,R~\simeq~\frac12\,\emetr\,\invemetr[\rho][\sigma]\Ricci[\rho][\sigma]
                 \=\frac12\,\emetr\,\left(\dmup[\rho]\dmup[\sigma]\hmetr[\rho][\sigma]-\dd^2h\right)\;,
\end{equation}
so that the l.h.s.\ of the Einstein equations in weak field approximation reads
\begin{equation}
\begin{split}
\GEinst&\=\Ricci-\dfrac12\,\gmetr\,R ~\simeq~\\
&~\simeq~\dmup[\rho]\dm[{(}\mu]\hmetr[\nu{)}][\rho]-\frac12\,\dd^2\hmetr-\frac12\,\dm\dm[\nu]h
-\frac12\,\emetr\left(\dmup[\rho]\dmup[\sigma]\hmetr[\rho][\sigma]-\dd^2h\right)\;.
\end{split}
\end{equation}
If one introduces the symmetric tensor
\begin{equation}
\bhmetr\=\hmetr-\frac12\,\emetr\,h\;,
\end{equation}
the above expression can be rewritten as
\begin{equation}
\begin{split}
\GEinst &~\simeq~
\dmup[\rho]\dm[{(}\mu]\bhmetr[\nu{)}][\rho]-\frac12\,\dd^2\bhmetr
-\frac12\,\emetr\,\dmup[\rho]\dmup[\sigma]\bhmetr[\rho][\sigma]\=\\
&\=\dmup[\rho]\dm[{[}\nu]\bhmetr[\rho{]}][\mu]+\dmup[\rho]\dmup[\sigma]\emetr[\mu][{[}\sigma]\,\bhmetr[\nu{]}][\rho]\=\\
&\=\dmup[\rho]\left(
          \dm[{[}\nu]\bhmetr[\rho{]}][\mu]+\dmup[\sigma]\emetr[\mu][{[}\rho]\,\bhmetr[\nu{]}][\sigma]
              \right)\;,
\end{split}
\end{equation}
where we have exchanged dumb indices in the last term of the second line.\par

If we now define the tensor
\begin{equation} \label{eq:Gscr}
\Gscr~\equiv~
\dm[{[}\nu]\bhmetr[\rho{]}][\mu]+\dmup[\sigma]\emetr[\mu][{[}\rho]\,\bhmetr[\nu{]}][\sigma]\;,
\end{equation}
whose structure implies the property
\begin{equation}
\Gscr\=-\Gscr[\mu][\rho][\nu]\;,
\end{equation}
the Einstein equations can be rewritten in the compact form:
\begin{equation}\label{eq:Einst}
\boxed{\;\GEinst\=\dmup[\rho]\Gscr\=8\pi\GN\;T_{\mu\nu}\,}\quad.
\end{equation}
We can impose a gauge fixing making use of the \emph{harmonic coordinate condition}, expressed by the relation \cite{Wald:1984rg}:
\begin{equation}
\dm\left(\sqrt{-g}\,\invgmetr\right)=0
\;\quad\Leftrightarrow\quad\;
\Box x^\mu=0\;,
\end{equation}
where $g\equiv\mathrm{det}\left[\gmetr\right]$, and that can be rewritten in the form
\begin{equation}
\invgmetr\,\Gam\,=\,0\;,
\end{equation}
also known as \emph{De Donder gauge}.\par
Imposing the above condition and using eqs.\ \eqref{eq:gmetr} and
\eqref{eq:Gam}, in first-order approximation we find:
\begin{equation}
0~\simeq~ \frac12\,\invemetr\,\invemetr[\lambda][\rho]\left(\dm[\mu]\hmetr[\nu][\rho]+\dm[\nu]\hmetr[\rho][\mu]-\dm[\rho]\hmetr[\mu][\nu]\right)\=
\dm\invhmetr-\frac12\,\dmup[\nu]h\;,
\end{equation}
that is, we have the condition
\begin{equation} \label{eq:gaugecond0}
\dm\invhmetr\simeq\frac12\,\dmup[\nu]h
\;\quad\Leftrightarrow\;\quad
\dmup\hmetr\simeq\frac12\,\dm[\nu]h\;\;.
\end{equation}
Now, one also has
\begin{equation}
\dmup\hmetr\=\dmup\left(\bhmetr+\frac12\,\emetr h\right)
\=\dmup\bhmetr+\frac12\,\dm[\nu]h\;,
\end{equation}
and, using eq.\ \eqref{eq:gaugecond0}, we find the so-called \emph{Lorenz gauge condition}:
\begin{equation}
\dmup\bhmetr~\simeq~0\;.
\end{equation}
The above relation further simplifies expression \eqref{eq:Gscr} for $\Gscr$, which takes the very simple form
\begingroup%
\setlength{\abovedisplayshortskip}{2pt plus 3pt}%
\setlength{\belowdisplayshortskip}{12pt plus 3pt minus 4pt}
\begin{equation}
\boxed{\;\Gscr~\simeq~\dm[{[}\nu]\bhmetr[\rho{]}][\mu]\,}\;\;,
\end{equation}
\endgroup
and verifies also the relation
\begin{equation}
\dm[{[}\lambda{|}]\Gscr[0][{|}\mu][\nu{]}]\=0
\;\qRq\;
\Gscr[0][\mu][\nu] \propto \dm\mathcal{A}_\nu-\dm[\nu]\mathcal{A}_\mu\;,
\end{equation}
which implies the existence of a potential.

\paragraph{Gravito-Maxwell equations.}
Now, let us define the fields%
\footnote{for the sake of simplicity, we initially set the physical charge $e=m=1$}%
\begin{subeqsds}{6}{12}\label{eq:fields0}
\begin{align}
\Eg&~\equiv~E_i~=\,-\,\frac12\,\Gscr[0][0][i]~=\,-\,\frac12\,\dm[{[}0]\bhmetr[i{]}][0]\;,\\[\jot]
\Ag&~\equiv~A_i\=\frac14\,\bhmetr[0][i]\;,\\[\jot]
\Bg&~\equiv~B_i
                 \=\frac14\,{\varepsilon_i}^{jk}\,\Gscr[0][j][k]\;,
\end{align}
\end{subeqsds}
where obviously $i=1,2,3$ and
\begin{equation}
\Gscr[0][i][j]\=\dm[{[}i]\bhmetr[j{]}][0]
\=\frac12\left(\dm[i]\bhmetr[j][0]-\dm[j]\bhmetr[i][0]\right)\=4\,\dm[{[}i]A_{j{]}}\;.
\end{equation}
One can immediately see that
\begin{equation}
\begin{split}
\Bg&\=\frac14\,{\varepsilon_i}^{jk}\,4\,\dm[{[}j]A_{k{]}}
   \={\varepsilon_i}^{jk}\,\dm[j]A_k=\nabla\times\Ag\;,\\[3\jot]
&~\Longrightarrow\quad \nabla\cdot\Bg\=0\;.
\end{split}
\end{equation}
Then one also has
\begin{equation}
\nabla\cdot\Eg\=\dmup[i]E_i\=-\dmup[i]\frac{\Gscr[0][0][i]}{2}
              \=-8\pi\GN\;\frac{T_{00}}{2}
              \=4\pi\GN\;\rhog\;,
\end{equation}
using eq.\ \eqref{eq:Einst} and having defined $\rhog\equiv-T_{00}$\,.\par\bigskip
If we consider the curl of $\Eg$, we obtain
\begin{equation}
\begin{split}
\nabla\times\Eg&\={\varepsilon_i}^{jk}\,\dm[j]E_k
               \=-{\varepsilon_i}^{jk}\,\dm[j]\frac{\Gscr[0][0][k]}{2}
               \=-\frac12\,{\varepsilon_i}^{jk}\,\dm[j]\dm[{[}0]\bhmetr[k{]}][0]\=\\[2\jot]
              &\=-\frac14\,4\;\dm[0]\,{\varepsilon_i}^{jk}\,\dm[j]A_k
               \=-\dm[0]B_i\=-\frac{\dd\Bg}{\dd t}\;.
\end{split}
\end{equation}
Finally, one finds for the curl of $\Bg$
\begin{equation}\label{eq:gravMaxwell4}
\begin{split}
\nabla\times\Bg&\={\varepsilon_i}^{jk}\,\dm[j]B_k
      \=\frac14\,{\varepsilon_i}^{jk}\,
                {\varepsilon_k}^{\ell m}\,\dm[j]\Gscr[0][\ell][m]\=\\[3\jot]
     &\=\frac14\left({\delta_i}^\ell\delta^{jm}-{\delta_i}^m\delta^{j\ell}\right)\dm[j]\Gscr[0][\ell][m]
      \=\frac12\,\dmup[j]\Gscr[0][i][j]\=\\[3\jot]
     &\=\frac12\left(\dmup\Gscr[0][i][\mu]+\dm[0]\Gscr[0][i][0]\right)
      \=\frac12\left(\dmup\Gscr[0][i][\mu]-\dm[0]\Gscr[0][0][i]\right)\=\\[3\jot]
     &\=\frac12\left(8\pi\GN\;T_{0i}-\dm[0]\Gscr[0][0][i]\right)
      \=4\pi\GN\;j_i+\frac{\dd E_i}{\dd t}\=\\[3\jot]
     &\=4\pi\GN\;\jg\+\frac{\dd\Eg}{\dd t}\;,
\end{split}
\end{equation}
using again eq.\ \eqref{eq:Einst} and having defined
$\jg \equiv j_i \equiv T_{0i}$\,.\par\smallskip
Summarizing, once defined the fields of \eqref{eq:fields0} and having restored physical units, one gets the field equations:
\begin{equation} \label{eq:gravMaxwell}
\begin{split}
\nabla\cdot\Eg&\=4\pi\GN\;\rhog\;;\\[2\jot]
\nabla\cdot\Bg&\=0 \;;\\[2\jot]
\nabla\times\Eg&~=-\dfrac{\dd\Bg}{\dd t} \;;\\[2\jot]
\nabla\times\Bg&\=\frac{4\pi\GN}{c^2}\;\jg
                  \+\frac{1}{c^2}\,\frac{\dd\Eg}{\dd t}\;,
\end{split}
\end{equation}
formally equivalent to Maxwell equations, where $\Eg$ and $\Bg$ are the gravitoelectric and gravitomagnetic field, respectively.
For example, on the Earth surface, $\Eg$ is simply the Newtonian gravitational acceleration and the $\Bg$ field is related to angular momentum interactions \cite{agop2000local,agop2000some,braginsky1977laboratory,huei1983calculation,peng1990new}.
The mass current density vector $\jg$ can also be expressed as:
\begin{equation}
\jg \= \rhog\,\mathbf{v} \;,
\end{equation}
where $\mathbf{v}$ is the velocity and $\rhog$ is the mass density.

\paragraph{Gravito-Lorentz force.}
Let us consider the geodesic equation for a particle in the field of a
weakly gravitating object:
\begin{equation}
\frac{d^2x^\lambda}{ds^2}\+\Gam\,\frac{dx^\mu}{ds}\,\frac{dx^\nu}{ds}\=0\;.
\end{equation}
If we consider a particle in non-relativistic motion, the velocity of the particle becomes $\frac{v_i}{c}\simeq\frac{dx^i}{dt}$. If we also neglect terms in the form $\frac{v_i\,v^j}{c^2}$ and limit ourselves to static fields $\left(\dm[t]\gmetr=0\right)$, it can easily be verified that a geodesic equation for a particle in non-relativistic motion can be written as \cite{ruggiero2002gravitomagnetic,Mashhoon:2003ax}:
\begin{equation}
\frac{d\mathbf{v}}{dt} \= \Eg+\mathbf{v}\times\Bg\;,
\end{equation}
which shows that the free fall of the particle is driven by the analogous of a Lorentz force produced by the gravito-Maxwell fields.

\paragraph{Generalized Maxwell equations.}
It is possible to define the generalized electric/magnetic field, scalar and vector potentials containing both electromagnetic and gravitational term as
\begin{equation} \label{eq:genpot}
\E=\Ee+\frac{m}{e}\,\Eg\,;\quad\;
\B=\Be+\frac{m}{e}\,\Bg\,;\quad\;
\phi=\phi_\textrm{e}+\frac{m}{e}\,\phig\,;\quad\;
\A=\Ae+\frac{m}{e}\,\Ag\,,
\end{equation}
where $m$ and $e$ are the mass and electronic charge, respectively, and the subscripts identify the electromagnetic and gravitational contributions.\par
The generalized Maxwell equations then become
\begin{equation} \label{eq:genMaxwell}
\begin{split}
\nabla\cdot\E&\=\left(\frac1\epsg+\frac{1}{\epsz}\right)\,\rho \;;\\[2\jot]
\nabla\cdot\B&\=0 \;;\\[2\jot]
\nabla\times\E&~=-\dfrac{\dd\B}{\dd t} \;;\\[2\jot]
\nabla\times\B&\=\left(\mug+\muz\right)\,\jj
                  \+\frac{1}{c^2}\,\dfrac{\dd\E}{\dd t} \;,
\end{split}
\end{equation}
where $\epsz$ and $\muz$ are the electric permittivity and magnetic permeability in the vacuum, and where we have set
\begingroup%
\begin{equation}
\begin{split}
\rhog&\=\frac{m}{e}\,\rho\;,\qquad\qquad\qquad\\[\jot]
\jg&\=\frac{m}{e}\,\jj\;,
\end{split}
\end{equation}
\endgroup
$\rho$ and $\jj$ being the electric charge density and electric current density, respectively. The introduced vacuum gravitational permittivity $\epsg$ and vacuum gravitational permeability $\mug$ are defined as
\begin{equation}
\epsg=\frac{1}{4\pi\GN}\,\frac{e^2}{m^2}\;,\qquad\quad
\mug=\frac{4\pi\GN}{c^2}\,\frac{m^2}{e^2}\;.
\end{equation}

\subsection{Generalizing London equations}
The London equations for a superfluid in stationary state read \cite{tinkham1996introduction,ketterson1999superconductivity,degennes1989superconductivity}:
\begin{subeqs} \label{eq:London}
\begin{align}
\Ee&\=\frac{m}{\ns\,e^2}\;\dfrac{\dd\jj}{\dd t}\;;\label{eq:London1} \\[2\jot]
\Be&~=-\frac{m}{\ns\,e^2}\;\nabla\times\jj \;.\label{eq:London2}
\end{align}
\end{subeqs}
where \;$\jj=\ns\,e\,\vs$\; is the supercurrent and \,$\ns$\, is the superelectron density.
If we also consider Ampère's law for a superconductor in stationary state (no displacement current)
\begin{equation} \label{eq:Ampere}
\nabla\times\Be\=\muz\,\jj\;,
\end{equation}
from \eqref{eq:London2} and using vector calculus identities, we obtain
\begin{equation}
\nabla\times\nabla\times\Be\=\nabla\left(\cancel{\nabla\cdot\Be}\right)-\nabla^2\Be
     \=\muz\,\nabla\times\jj~=-\muz\,\frac{\ns\,e^2}{m}\;\Be\;,
\end{equation}
that is,
\begin{equation}
\nabla^2\Be\=\frac{1}{\lambdae^2}\;\Be\;,
\end{equation}
where we have introduced the penetration depth
\begin{equation} \label{eq:lambdae}
\lambdae\=\sqrt{\frac{m}{\muz\,\ns\,e^2}}\;.
\end{equation}
Using the vector potential $\Ae$, the two London equations \eqref{eq:London} can be summarized in the (not gauge-invariant) form
\begin{equation} \label{eq:Londonsumm}
\qquad \jj~=-\frac{1}{\muz\,\lambdae^2}\;\Ae \qquad\qquad
\big(\,\Be\,=\,\nabla\times\Ae\,\big)\quad.
\end{equation}

\paragraph{Generalized London equations.}
If we now take into account gravitational corrections, we should consider for the fields and the vector potential the generalized form of definition \eqref{eq:genpot}:
\begin{equation}
\B=\Be+\frac{m}{e}\,\Bg\;,\qquad
\A=\Ae+\frac{m}{e}\,\Ag\;,\qquad
\B=\nabla\times\A\;.
\end{equation}
If $\A$ is minimally coupled to the wave function
\begin{equation} \label{eq:psi}
\qquad \psi\=\psi_0\,\e^{i\varphi}\;, \qquad\qquad \psi_0^2\equiv\abs{\psi}^2=n_s\;,
\end{equation}
the second London equation can be derived from a quantum mechanical current density
\begin{equationds}{12}{15}
\jj ~= -\frac{i}{2m}\left(\psi^\ast\tilde{\nabla}\psi-\psi\tilde{\nabla}\psi^\ast\right)\;,
\end{equationds}
where $\tilde{\nabla}$ is the covariant derivative for the minimal coupling:
\begin{equation}
\tilde{\nabla}\=\nabla-i\,\tilde{g}\,\A\;,
\end{equation}
so that one has for the current
\begin{equation}
\jj ~=
\,-\frac{i}{2m}\left(\psi^\ast\nabla\psi-\psi\nabla\psi^\ast\right)-\frac{\tilde{g}}{m}\,\A\,\abs{\psi}^2
\= \frac{1}{m}\,\abs{\psi}^2\left(\nabla\varphi-\tilde{g}\,\A\right)\;.
\end{equation}
If we now take the curl of the previous equation, we find
\begin{equation}
\B~=\,-\,\frac{m}{\tilde{g}\,\abs{\psi}^2}\;\nabla\times\jj
   \= -\,\frac1\zeta\;\nabla\times\jj\;,
\end{equation}
which is the generalized form of the second London equation \eqref{eq:London2}.\par\medskip
To find an explicit expression for $\zeta$, we consider the case $\Bg=0$ obtaining
\begin{equation}
\B \= \Be+\frac{m}{e}\,\xcancel{\Bg} ~= -\,\frac1\zeta\;\nabla\times\jj\;,
\end{equation}
and, using \eqref{eq:London2}, \eqref{eq:lambdae} and \eqref{eq:psi}, we find
\begin{equation}
\tilde{g}\,=\,e^2\;, \qquad\quad \frac1\zeta\=\muz\,\lambdae^2\;.\qquad
\end{equation}
Then we consider the case $\Be=0$, so that we have
\begin{equation}
\B \= \xcancel{\Be}+\frac{m}{e}\,\Bg ~= \,-\muz\,\lambdae^2\;\nabla\times\jj
   ~= \,-\muz\,\lambdae^2\;\frac{m}{e}\,\nabla\times\jg\;,
\end{equation}
together with gravito-Ampère's law \eqref{eq:gravMaxwell} in stationary state,
\begin{equation}
\nabla\times\Bg\=\mug\;\jg\;,
\end{equation}
so that, taking the curl of the above equation, we find
\begin{equation}
\nabla\times\nabla\times\Bg~=\,-\nabla^2\Bg
     \=\mug\,\nabla\times\jg~=-\,\mug\,\frac{1}{\muz\,\lambdae^2}\;\Bg
     ~=\,-\,\frac{1}{\lambdag^2}\;\Bg\;,
\end{equation}
where we have introduced the penetration depth
\begin{equation} \label{eq:lambdag}
\lambdag\=\sqrt{\frac{\muz\,\lambdae^2}{\mug}}
        \=\sqrt{\frac{c^2}{4\pi\GN\,m\,n_s}}\;.
\end{equation}
Finally, using the stationary generalized Ampère's law from \eqref{eq:genMaxwell} and using eq.\ \eqref{eq:lambdag} we find
\begin{equation} \label{eq:genAmpere}
\nabla\times\B\=\left(\muz+\mug\right)\,\jj\=\muz\left(1+\frac{\lambdae^2}{\lambdag^2}\right)\jj\;,
\end{equation}
and taking the curl we obtain the general form
\begin{equation}
\begin{split}
\nabla^2\B&~=\,-\muz\left(1+\frac{\lambdae^2}{\lambdag^2}\right)\,\nabla\times\jj
\=\muz\,\frac{1}{\muz\,\lambdae^2}\,\left(1+\frac{\lambdae^2}{\lambdag^2}\right)\,\B\=\\[\jot]
&\=\left(\frac{1}{\lambdae^2}+\frac{1}{\lambdag^2}\right)\,\B
\=\frac{1}{\lambda^2}\;\B\;,
\end{split}
\end{equation}
where we have defined a \emph{generalized penetration depth} $\lambda$\,:
\begin{equation}
\lambda\=\frac{\lambdag\,\lambdae}{\sqrt{\lambdag^2+\lambdae^2}}
       ~\simeq~ \lambdae \qquad\quad \left(\frac{\lambdag}{\lambdae}\simeq{10}^{21}\right)\;\;.
\end{equation}
The general form of eq.\ \eqref{eq:Londonsumm} is
\begin{equation}
\qquad \jj~=-\,\zeta\,\A \qquad\qquad
\big(\,\B\,=\,\nabla\times\A\,\big)\;\;.
\end{equation}
and, since charge-conservation requires the condition ${\nabla\cdot\jj=0}$, we obtain for the vector potential
\begin{equation*}
\nabla\cdot\A=0 \;,
\end{equation*}
that is, the so-called \emph{Coulomb gauge} (or \emph{London gauge}).

\pagebreak

\section{Isotropic superconductor} \label{sec: Isotropic SC}
In Sect.\ \ref{sec:Intro} we have shown how Modanese was able to theoretically describe the gravitational shielding effect due to the presence of a superfluid. Now we are going to study the same problem with a different approach.\par
Modanese has solved gravitational field equation where the contribution of the superfluid was encoded in the energy-momentum tensor. In the following, we are going to solve the Ginzburg-Landau equation for the superfluid order parameter in an external gravitational field.\par
Let us restrict ourselves to the case of an isotropic superconductor in the gravitational field of the earth and in absence of an electromagnetic field, we can take $\Ee=0$ and $\Be=0$.
Moreover, $\Bg$ in the solar system is very small \cite{mashhoon1989detection,ljubivcic1992proposed}, therefore \,$\E=\frac{m}{e}\,\Eg$\, and \,$\B=0$. Finally, we also have the relations \,$\phi=\frac{m}{e}\,\phig$\, and \,$\A=\frac{m}{e}\,\Ag$\,, so we can write down our set of conditions:
\begin{subeqs} \label{eq:fields}
\begin{gather}
\Ee=0\;, \;\quad  \Be=0\;, \;\quad  \Bg=0
\;\quad\Longrightarrow\quad\;
\E=\frac{m}{e}\;\Eg\;, \;\quad  \B=0\;;\;\;
\intertext{together with}
\phi=\frac{m}{e}\,\phig \;,\qquad\quad
\A=\frac{m}{e}\,\Ag\;.
\end{gather}
\end{subeqs}
The situation is not the same as the Meissner effect but, rather, as the case of a superconductor in an electric field.

\subsection{Time-dependent Ginzburg-Landau equations}
Since the gravitoelectric field is formally analogous to an electric field we can use the time-dependent Ginzburg-Landau equations (TDGL) which, in the Coulomb gauge $\nabla\cdot\A=0$ are written in the form \cite{tang1995time,lin1997ginzburg,ullah1991effect,ghinovker1999explosive,kopnin1999time,fleckinger1998dynamics,du1996high}:
\begin{subeqsds}{12}{6} \label{eq:TDGL}
\begin{gather}
\frac{\hbar^2}{2\,m\,\DD}\left(\frac{\dd}{\dd t}
+\frac{2\,i\,e}{\hbar}\,\phi\right)\,\psi
\-a\,\psi\+b\,\abs{\psi}^2\psi\+\frac{1}{2\,m}\left(i\hbar\nabla
+\frac{2\,e}{c}\,\A\right)^2 \psi\=0 \;,\label{eq:TDGLn1}\\[2\jot]
\nabla\times\nabla\times\A-\nabla\times\HH~=-\frac{4\pi}{c}\,
\left(\frac{\sigma}{c}\,\frac{\dd\A}{\dd t}+\sigma\,\nabla\phi
+\frac{i\hbar\,e}{m}\left(\psi^*\nabla\psi-\psi\nabla\psi^*\right)
+\frac{4\,e^2}{mc}\,\abs{\psi}^2\A\right)\,,
\end{gather}
\end{subeqsds}
where $\DD$ is the diffusion coefficient, $\sigma$ is the conductivity in the normal phase, $\HH$ is the applied field and the vector field $\A$ is minimally coupled to $\psi$. The above TDGL equations for the variables $\psi$, $\A$ are derived minimizing the total Gibbs free energy of the system \cite{tinkham1996introduction,ketterson1999superconductivity,degennes1989superconductivity}.\par\smallskip
The coefficients $a$ and $b$ in \eqref{eq:TDGLn1} have the following form:
\begin{equation}
\begin{split}
a&\=a(T)\=a_\ms{0}\,(T-\Tc)\;,\\[\jot]
b&\=b(T)~\equiv~b(\Tc)\;,
\end{split}
\end{equation}
$a_\ms{0\,}$, $b$ being positive constants and \Tct the critical temperature of the superconductor. The boundary and initial conditions are
\begingroup%
\setlength{\belowdisplayskip}{7pt plus 3pt minus 4pt}%
\begin{align}
  \left.
  \begin{aligned}
  \left(i\hbar\,\nabla\psi+\dfrac{2\,e}{c}\,\A\,\psi\right)\cdot\n=0&  \cr
  \hfill\nabla\times\A\cdot\n=\HH\cdot\n&  \\[1.5\jot]
  \hfill\A\cdot\n=0&
  \end{aligned}
  \;\;\right\} \; \text{on }\dd\Omega\times(0,t)\;;
\qquad
  \left.
  \begin{aligned}
  \psi(x,0)&\=\psi_0(x) \cr
  \A(x,0)&\=\A_0(x)
  \end{aligned}
  \!\!\!\!\right\} \; \text{on }\Omega\qquad
\label{eq:boundary}
\end{align}
\endgroup
where $\dd\Omega$ is the boundary of a smooth and simply connected domain in $\mathbb{R}^\ms{\textrm{N}}$.\par
\paragraph{Dimensionless TDGL.}
In order to write eqs.\ \eqref{eq:TDGL} in a
dimensionless form, the following quantities can be introduced:
\begin{subeqs} \label{eq:param}
\begin{gather}
\Psi^2(T)\=\frac{\abs{a(T)}}{b}\;,\qquad
\xi(T)\=\frac{h}{\sqrt{2\,m\,|a(T)|}}\;,\qquad
\lambda(T)\=\sqrt{\frac{b\,m\,c^2}{4\pi\,|a(T)|\,e^2}}\;,\\[1.2\jot]
\Hc(T)\=\sqrt{\frac{4\pi\,\muz\,\abs{a(T)}^2}{b}}\=
\frac{h}{4\,e\,\sqrt{2\pi}\,\lambda(T)\,\xi(T)}\;,\\[5\jot]
\kappa\=\frac{\lambda(T)}{\xi(T)}\;,\qquad
\tau(T)\=\frac{\lambda^2(T)}{\DD}\;,\qquad
\eta\=\frac{4\pi\,\sigma\,\DD}{\epsz\,c^2}\;,
\end{gather}
\end{subeqs}
where $\lambda(T)$, $\xi(T)$ and $\Hc(T)$ are the penetration depth, coherence length and thermodynamic field, respectively. The dimensionless quantities are then defined as:%
\begin{subeqsds}{2}{15} \label{eq:dimlessfields}
\begin{gather}
x'=~\frac{x}{\lambda}\;,\qquad
t'=~\frac{t}{\tau}\;,\qquad
\psi'=~\frac{\psi}{\Psi}\;,
\intertext{and the dimensionless fields are written \bigskip}
\A'\=\frac{\A\,\kappa}{\sqrt{2}\,\Hc\,\lambda}\;,\qquad
\phi'\=\frac{\phi\,\kappa}{\sqrt{2}\,\Hc\,\DD}\;,\qquad
\HH'\=\frac{\HH\,\kappa}{\sqrt{2}\,\Hc}\;.
\end{gather}
\end{subeqsds}
Inserting eqs.~(\ref{eq:dimlessfields}) in eqs.~\eqref{eq:TDGL} and dropping the prime gives the dimensionless TDGL equations in a bounded, smooth and simply connected domain in $\mathbb{R}^\ms{\textrm{N}}$ \cite{tang1995time,lin1997ginzburg}:
\begin{subeqs}\label{eq:dimlessTDGL}
\begin{gather}
\frac{\dd\psi}{\dd t}\+i\,\phi\,\psi
\+\kappa^2\left(\abs{\psi}^2-1\right)\,\psi
\+\left(i\,\nabla+\A\right)^2 \psi\=0 \;,\label{subeq:dimlessTDGLn1} \\[2\jot]
\nabla\times\nabla\times\A-\nabla\times\HH~=
-\eta\,\left(\frac{\dd\A}{\dd t}+\nabla\phi\right)
-\frac{i}{2}\left(\psi^*\nabla\psi-\psi\nabla\psi^*\right)
-\abs{\psi}^2\A\,,\label{subeq:dimlessTDGLn2}
\end{gather}
\end{subeqs}
and the boundary and initial conditions~\eqref{eq:boundary} become, in the dimensionless form
\begin{align}
  \left.
  \begin{aligned}
  \left(i\,\nabla\psi+\A\,\psi\right)\cdot\n=0&\cr
  \nabla\times\A\cdot\n=\HH\cdot\n&\cr
  \A\cdot\n=0&
  \end{aligned}
  \!\!\!\right\} \; \text{on }\dd\Omega\times(0,t)\;;
\qquad
  \left.
  \begin{aligned}
  \psi(x,0)&\=\psi_0(x)\cr
  \A(x,0)&\=\A_0(x)
  \end{aligned}
  \!\!\!\!\right\} \; \text{on }\Omega\;.\qquad
\label{eq:dimlessboundary}
\end{align}

\subsection{Solving dimensionless TDGL}
If the superconductor is on the Earth's surface, the gravitational field is very weak and approximately constant. This means that one can write
\begin{equation}
\phi=-\gstar\,x\;,
\end{equation}
with
\begingroup%
\setlength{\abovedisplayshortskip}{7pt plus 3pt}%
\setlength{\belowdisplayskip}{15pt plus 3pt minus 9pt}%
\setlength{\belowdisplayshortskip}{15pt plus 3pt minus 4pt}
\begin{equation}
\gstar\=\frac{\lambda(T)\,\kappa\,m\,g}{\sqrt{2}\,e\,\Hc(T)\,\DD}~\ll~1\;,
\end{equation}
\endgroup
$g$ being the acceleration of gravity. The corrections to $\phi$ in the superconductor are of second order in $\gstar$ and therefore they are not considered here.\par\medskip
Now we search for a solution of the form
\begin{equation} \label{eq:condfield}
\begin{split}
\psi(x,t)&\=\psi_0(x,t)+\gstar\,\gamma(x,t)\;,\\
A(x,t)&\=\gstar\,\beta(x,t)\;,\\
\phi(x)&~=-\gstar\,x\;.
\end{split}
\end{equation}
At order zero in $\gstar$, eq.~(\ref{subeq:dimlessTDGLn1}) gives
\begin{equation} \label{eq:dimlessTDGL0}
\frac{\dd\psi_0(x,t)}{\dd t} \+ \kappa^2\left(\abs{\psi_0(x,t)}^2-1\right)\,\psi_0(x,t)
\- \frac{\dd^2\psi_0(x,t)}{\dd x^2} \=0 \;,
\end{equation}
with the conditions
\begin{equation}
\begin{split}
\psi_0(x,0)&\=0\;,\\
\psi_0(0,t)&\=0\;,\\
\psi_0(L,t)&\=0\;,
\end{split}
\end{equation}
where $L$ is the length of the superconductor, here in units of $\lambda$, and $t=0$ is the instant in which the material undergoes the transition to the superconducting state.\par
The static classical solution of eq.~\eqref{eq:dimlessTDGL0} is
\begin{equation} \label{eq:psi0}
\psi_0(x,t)~\equiv~\psi_0(x)\=\tanh\left(\frac{\kappa\,x}{\sqrt{2}}\right)\,\tanh\left(\frac{\kappa\,\left(x-L\right)}{\sqrt{2}}\right)\;,
\end{equation}
and, from \eqref{subeq:dimlessTDGLn1}, one obtains
\begin{equation} \label{eq:gamma}
\frac{\dd\gamma(x,t)}{\dd t} \- \frac{\dd^2\gamma(x,t)}{\dd x^2}\+ \kappa^2\left(3\,\abs{\psi_0(x)}^2-1\right)\,\gamma(x,t)\=
i\,x\,\psi_0(x)
\end{equation}
at first-order in $\gstar$, with the conditions
\begin{equation}
\begin{split}
\gamma(x,0)&\=0\;,\\
\gamma(0,t)&\=0\;,\\
\gamma(L,t)&\=0\;.
\end{split}
\end{equation}
The first-order equation for the vector potential is written
\begin{equation} \label{eq:vectpot}
\eta\,\frac{\dd\beta(x,t)}{\dd t} \+ \abs{\psi_0(x)}^2\,\beta(x,t)
\+ J(x,t) \- \eta \= 0 \;,
\end{equation}
with the constraint
\begin{equation} \label{eq:betanull}
\beta(x,0) \= 0 \;.
\end{equation}
The second-order spatial derivative of $\beta$ does not appear in eq.~\eqref{eq:vectpot}: this is due to the fact that, in one
dimension, one has
\begin{equation}
\nabla^2 A \= \frac{\dd}{\dd x}\,\nabla\cdot\A \;,
\end{equation}
and therefore, in the Coulomb gauge
\begin{equation}
\nabla\times\nabla\times\A \=
\nabla\left(\nabla\cdot\A\right) - \nabla^2 A \= 0 \;.
\end{equation}
The quantity $J(x,t)$ that appears in eq.~\eqref{eq:vectpot} is given by
\begin{equation} \label{eq:J}
J(x,t) \= \frac12 \left(
\psi_0(x)\,\frac{\dd}{\dd x}\,\textrm{Im}\left[\gamma(x,t)\right]
-\textrm{Im}\left[\gamma(x,t)\right]\,\,\frac{\dd}{\dd x}\psi_0\right)\;,
\end{equation}
and the solution of eq.~\eqref{eq:vectpot} is
\begin{align} \label{eq:beta}
\beta(x,t) &\= \frac1\PP \left(1-\e^{-\PP\,t}\right) \-
\frac{\e^{-\PP\,t}}{\eta}\,\int^{t}_\ms{0} dt\,J(x,t)\;\e^{\PP\,t} \;\;,
\intertext{with}
&~ \PP \= \frac{\abs{\psi_0(x)}^2}{\eta}\;.
\end{align}
Now, we have the form~\eqref{eq:psi0} for $\psi_0(x,t)$ and also the above~\eqref{eq:beta} for $\beta(x,t)$ as a function of $\gamma(x,t)$ through the definition of $J(x,t)$: the latter can be used in \eqref{eq:condfield} to obtain both $\psi(x,t)$ and $\A(x,t)$ as functions of $\gamma(x,t)$.\par\smallskip
The gravitoelectric field can be found using the relation
\begin{equation} \label{eq:EgphiA}
\Eg~= -\,\nabla\phi - \frac{\dd\A}{\dd t}\;,
\end{equation}
and its explicit form reads
\begin{equation} \label{eq:Eg}
\frac1{\gstar}\;{\Eg(x,t)} \= 1 \- \e^{-\PP\,t} \-
\frac{\dd}{\dd t} \left(
\frac{\e^{-\PP\,t}}{\eta}\,\int^{t}_\ms{0} dt\,J(x,t)\;\e^{\PP\,t}\right)\;.
\end{equation}
The above formula shows that, for maximizing the effect of the reduction of the gravitational field in a superconductor, it is necessary to reduce $\eta$ and have large spatial derivatives of $\psi_0(x)$ and $\gamma(x,t)$. The condition for a small value of $\eta$ is a large normal-state resistivity for the superconductor and a small diffusion coefficient
\begin{equation}
\DD ~\sim~ \frac{\vF\,\ell}{3}\;,
\end{equation}
where $\vF$ is the Fermi velocity (which is small in HTSC) and \,$\ell$\, is the mean free path: this means that the effect is enhanced in "bad" samples with impurities, not in single crystals.\par\smallskip
If we consider the case $J(x,t)=0$, given by the condition
\begin{equation}
\psi_0(x)\=\textrm{Im}\left[\gamma(x,t)\right]~\equiv~\textrm{Im}\left[\gamma(x)\right]\;,
\end{equation}
we obtain the simplified equation
\begin{equation}
\eta\,\frac{\dd\beta(x,t)}{\dd t} \+\abs{\psi_0(x)}^2\,\beta(x,t)\-\eta\=0\;,
\end{equation}
which is solved, together with the constraint \eqref{eq:betanull}, by the function
\begin{equation}
\beta(x,t)\=\frac{\eta}{\abs{\psi_0(x)}^2}\,\left(1-\e^\mathlarger{-\frac{\abs{\psi_0(x)}^2}{\eta}\,t}\right)\;.
\end{equation}
Using then eqs.\ \eqref{eq:EgphiA} and \eqref{eq:condfield} we find
\begin{equation}
\frac{\Eg}{\gstar}\=1-\e^\mathlarger{-\frac{\abs{\psi_0(x)}^2}{\eta}\,t}\;.
\end{equation}
The above equation shows that, unlike the general case, in the absence of the contribution of $J(x,t)$ the effect is bigger than in the case of single crystal low-\Tct superconductor, where $\eta$ is large.

\subsubsection{Approximate solution}
From the experimental viewpoint, the greater are the length and time scales over which there is a variation of $\Eg$, the easier is the observation of this effect. Actually, we started from dimensionless equations and therefore the length and time scales are determined by $\lambda(T)$ and $\tau(T)$ of eqs.~\eqref{eq:param}, which should therefore be as large as possible. In
this sense, materials having very large $\lambda(T)$ could be interesting for the study of this effect \cite{blackstead2000magnetism}. Moreover, eq.~\eqref{eq:Eg} shows the dependence of relaxation with respect to $\abs{\psi_0(x)}^2$ through the definition of $\PP$: one can see that $\abs{\psi_0(x)}$ must be as small as possible and this implies that also $\kappa$ must be small, see eq.~\eqref{eq:psi0}. This also means that $\lambda(T)$ and $\xi(T)$ must both be large.\par\smallskip
Up to now we have dealt with the expression of $\beta(x,t)$ as a function of $\gamma(x,t)$. If we want to obtain an explicit expression for $\Eg$, we have to solve the equation~\eqref{eq:gamma} for $\gamma(x,t)$: this is a difficult task which can be undertaken only in a numerical way. Nevertheless, if one puts $\psi_0(x)\approx1$, which is a good approximation in the case of YBa${}_2$Cu${}_3$O${}_7$ (YBCO) in which $\kappa=94.4$, one can find the simple approximate solution:
\begin{equation} \label{eq:gammasol}
\gamma(x,t) \= i\,\gamma_0(x)
\+ i\,\sum^\infty_{n=1} \Qn \, \sin\left(\wn x\right) \,
\e^{-\Cnq\,t}\;,
\end{equation}
with
\begin{subeqs}
\begin{align}
\gamma_0(x) &\= \frac{x}{2\,\kappa^2}\,\Bigg(1-\ch[\frac{\alpha}{2}-\frac{\alpha}{L}\,x]\,\sech[\frac{\alpha}{2}]\Bigg)\;, \\[4\jot]
\Qn &\= \frac1L\,
\int^{\ms{L}}_\ms{0} dx\:\gamma_0(x)\,\sin(\wn x)
\= \frac{(-1)^n}{2\,\kappa^2}\,\Biggl(
\frac{1}{\wn}\-\wn\,\frac{Q_{n}^{{}^{(1)}}+Q_{n}^{{}^{(2)}}}{\Cnq}\Biggr)\;,\\[-5\jot]
\intertext{and}
\Cnq&\=\wn^2+2\,\kappa^2\;,\qquad\wn\=n\,\pi/L\;,\qquad\alpha\=\sqrt{2}\,\kappa\,L\;,\\[3\jot]
Q_{n}^{{}^{(1)}} &\= (-1)^n-\cosh\alpha
      \+\frac{2\,\alpha\,\wn}{L\,\Cnq}\,\sinh\alpha\;,\\[2\jot]
Q_{n}^{{}^{(2)}} &\=
   \left(\cosh\alpha-1\right)\,\Bigg(1 \+ \frac{2\,\alpha}{L^2\,\Cnq}\,
   \frac{(-1)^n-\cosh\alpha}{\sinh\alpha}\Bigg)\;.
\end{align}
\end{subeqs}
Taking into account eq.~\eqref{eq:psi0} and inserting eq.~\eqref{eq:gammasol} in eq.~\eqref{eq:J} and then in eq.~\eqref{eq:beta}, we can find a new expression for the gravitoelectric field $\Eg$:
\begin{equation} \label{eq:Egsol}
\frac1{\gstar}\;{\Eg(x,t)} \=
       1 \- \e^{-\PP\,t}\,\left(1-\frac{J_0(x)}{\eta}\right)
       \+ \frac{1}{\eta}\:\sum^\infty_{n=1}\,\Qn\:\Rn(x)\:\Sn(x,t) \;,
\end{equation}
where
\begin{subeqs}
\begin{align}
J_0(x) &\= \frac{1}{2\,\kappa^2}\,\left(
\psi_0(x)\,\frac{\dd}{\dd x}\gamma_0(x)
-\gamma_0(x)\,\frac{\dd}{\dd x}\psi_0(x)\right)\;, \\[4\jot]
\Rn(x) &\= \wn\,\psi_0(x)\,\cos(\wn x)
            \- \sin(\wn x)\frac{\dd}{\dd x}\psi_0(x)\;,\\[4\jot]
\Sn(x,t)&\=\frac{\Cnq\;\e^{-\Cnq\,t}\-\PP\,\e^{-\PP\,t}}{\PP-\Cnq}\;.
\end{align}
\end{subeqs}
By making the approximation
\begin{equation} \label{eq:drasticappr}
\gamma(x)~\simeq~\frac{i\,x}{2\,\kappa^2}\;,
\end{equation}
one finds the result
\begin{equation} \label{eq:Egappr}
\frac1{\gstar}\;{\Eg(x,t)} \=
    1 \- \e^{-\PP\,t}\,\left(1-\frac{J_{00}(x)}{\eta}\right)\;,
\end{equation}
where
\begingroup%
\setlength{\belowdisplayskip}{15pt plus 3pt minus 4pt}
\setlength{\belowdisplayshortskip}{15pt plus 3pt minus 4pt}
\begin{equation}
J_{00}(x) \= \frac{1}{2\,\kappa^2}\,
              \left(\psi_0(x)-x\,\frac{\dd}{\dd x}\psi_0(x)\right)\;.
\end{equation}
\endgroup
In spite of its crudeness, in the case of YBCO the above approximate solution~\eqref{eq:Egappr} gives the same results of the solution~\eqref{eq:Egsol}. Moreover, nothing changes significantly if one neglects the finite size of the superconductor and uses
\begin{equation}
\psi_0(x)\=\tanh\left(\kappa x/\sqrt2\right)
\end{equation}
instead of eq.~\eqref{eq:psi0}.

\subsection{YBCO vs.\ Pb}
\begin{figure}[!htp]
\captionsetup{skip=0pt,belowskip=15pt}
\centering
\includegraphics[width=\textwidth]{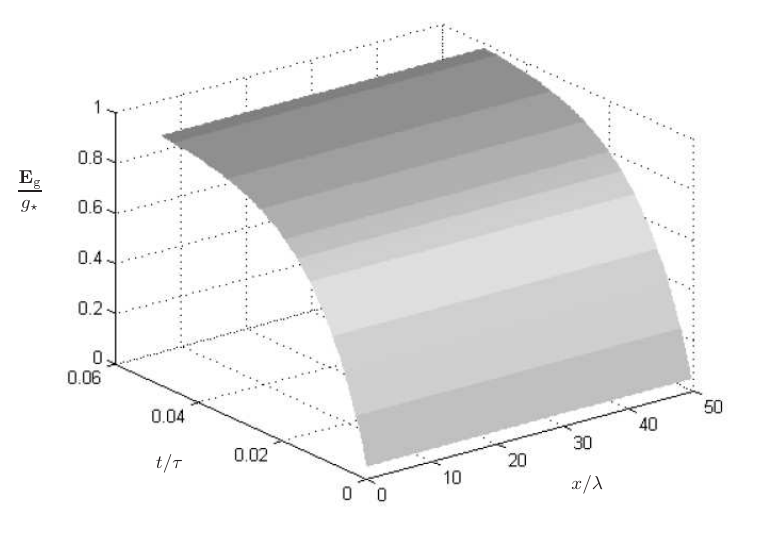}
\caption{The gravitational field $\Eg/\gstar$ as a function of the normalized time and space for YBCO at $T=77\,\K$}
\label{fig:YBCO}
\end{figure}

\begin{figure}[!htp]
\captionsetup{skip=-7pt}
\centering
\includegraphics[width=\textwidth]{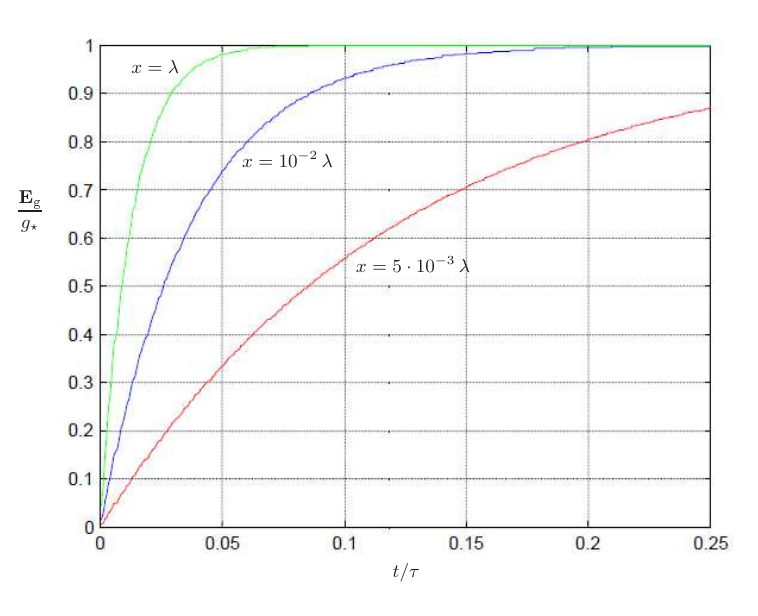}
\caption{The gravitational field as a function of the normalized time for increasing values of the $x$ variable for YBCO}
\label{fig:YBCOx}
\end{figure}

In the case of YBCO, the variation of the gravitoelectric field $\Eg$ in time and space is shown in Figs.~\ref{fig:YBCO} and \ref{fig:YBCOx}. It is easily seen that this effect is almost independent on the spatial coordinate.\par

\begin{figure}[!htp]
\captionsetup{skip=0pt,font=small,labelfont=small,format=hang}%
\centering
\includegraphics[width=\textwidth]{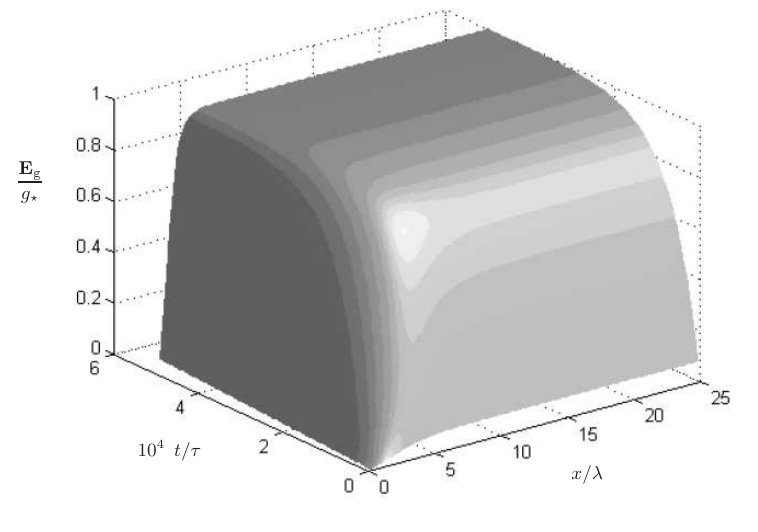}
\caption{The gravitational field $\Eg/\gstar$ as a function of the normalized time and space for Pb at $T=6.3\,\K$}
\label{fig:Pb}
\end{figure}

\begin{figure}[!htp]
\captionsetup{skip=-5pt,belowskip=5pt}%
\centering
\includegraphics[width=\textwidth]{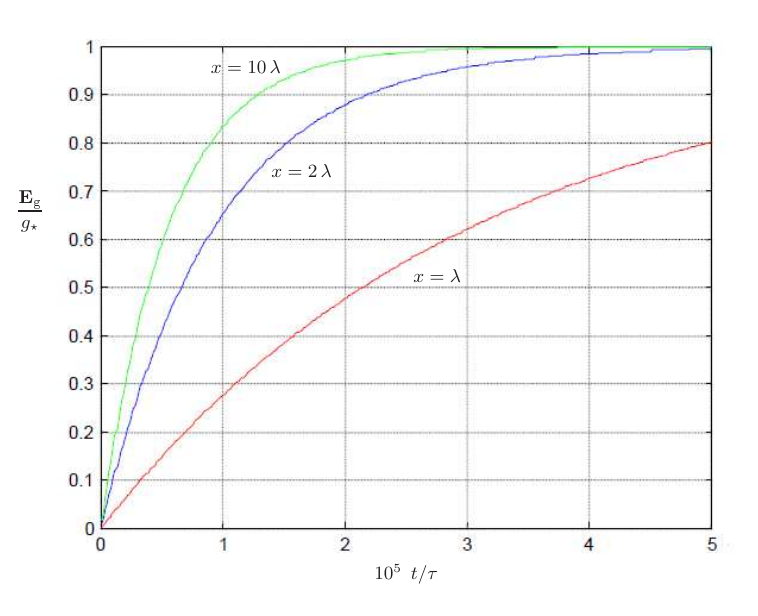}
\caption{The gravitational field as a function of the normalized time for increasing values of the $x$ variable for Pb}
\label{fig:Pbx}
\end{figure}

\begin{figure}[!htp]
\captionsetup{skip=-5pt}
\centering
\includegraphics[width=\textwidth]{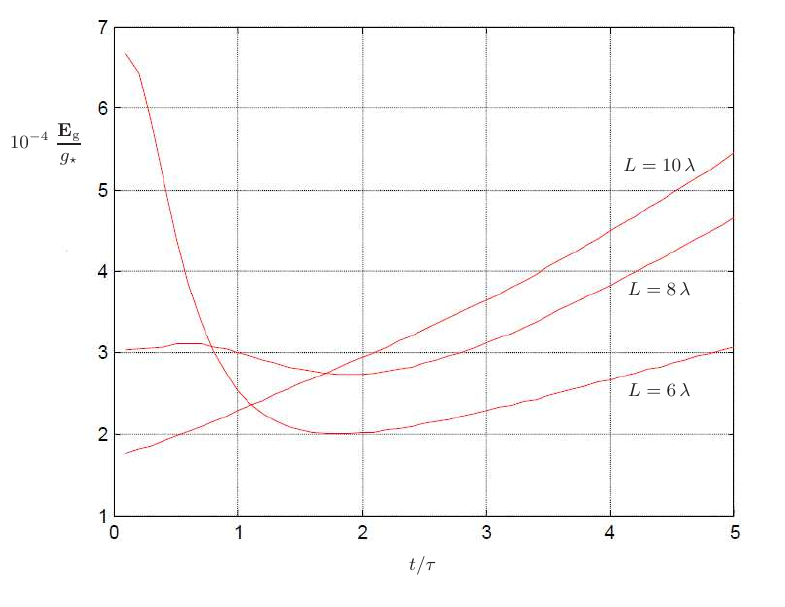}
\caption{The gravitational field $\Eg/\gstar$ as a function of the normalized time in the case of Pb, for different values of $L$ and $x=4\,\lambda$. The maximum of the shielding effect is evident.}
\label{fig:PbL}
\end{figure}

The results in the case of Pb are reported in Figs.~\ref{fig:Pb} and \ref{fig:Pbx}, which clearly show that, due to the very small value of $\kappa$, the reduction is greater near the surface. Moreover, in this particular case, some approximations made in the case of YBCO are no longer allowed: for example, the simplified relation~\eqref{eq:drasticappr} is not valid for small values of $L$. In fact, when $\kappa$ is small, the length $L$ plays an important role and, in particular, if $L$ is small the effect is remarkably enhanced, as shown in Fig.~\ref{fig:PbL}. In the same condition, a maximum of the effect (and therefore a minimum of $\Eg$) can occur at $t\neq0$, as can be seen in the same figure. In the extreme case $L=6\,\lambda$, we found that the system returns to the unperturbed value after a time \,$t_0\simeq10^5\,\tau$.\par
Table~\ref{tab:YBCOvsPb1} reports the values of the parameters of YBCO and Pb, calculated at a temperature $T_\star$ such that the quantity $\frac{T_\star-\Tc}{\Tc}$ is the same in the two materials. In Tables~\subref{subtab:YBCO} and \subref{subtab:Pb} are shown the calculated values of $\lambda$, $\tau$ and $\gstar$ at different temperatures.

\begin{table}[!htp]
\noindent\centering
\makegapedcells
\setcellgapes{5pt} 
\begin{tabular}
{@{} M{p}{0.175} M{p}{0.2} M{p}{0.275} @{}}
\toprule
\midrule
                   &  \text{YBCO}                 &  \text{Pb}  \\
\thinrule
\Tc                &    89~\K                     &  7.2~\K \\
T_\star            &    77~\K                     &  6.3~\K \\
\xi(T_\star)       &    3.6\cdot10^{-9}~\mt       &  1.7\cdot10^{-7}~\mt \\
\lambda(T_\star)   &    3.3\cdot10^{-7}~\mt       &  7.8\cdot10^{-8}~\mt \\
\makecell[t]{\sigma^{-1}}
                   &
                \makecell[t]{4\cdot10^{-7}~\Omega\,\mt\\[-2pt]
                        \text{\fns[{\ssmall\,$T\,=\,90\,\K$\,}]}}
                                                  &
                                               \makecell[t]{2.5\cdot10^{-9}~\Omega\,\mt\\[-2.pt]
                                                  \text{\fns[{\ssmall\,$T\,=\,15\,\K$\,}]}}\\
\Hc(T_\star)       &   0.2~\textrm{Tesla}          &  0.018~\textrm{Tesla}\\
\kappa             &   94.4                        &  0.48 \\
\tau(T_\star)      &   3.4\cdot10^{-10}~\s         &  6.1\cdot10^{-15}~\s \\
\eta               &   1.27\cdot10^{-2}            &  6.6\cdot10^{3} \\
\DD                &   3.2\cdot10^{-4}~\mt^{2}/\s  &  1~\mt^{2}/\s \\
\ell               &   6\cdot10^{-9}~\mt           &  1.7\cdot10^{-6}~\mt \\
\vF                &   1.6\cdot10^{5}~\mt/\s       &  1.83\cdot10^{6}~\mt/\s \\
\midrule
\bottomrule
\end{tabular}
\caption{YBCO vs.\ Pb}
\label{tab:YBCOvsPb1}
\end{table}


\vspace{0pt}
\begin{table}[!htp]
\small
\noindent
\centering
\makegapedcells
\setcellgapes{3pt}
\begin{adjustwidth}{-5cm}{-5cm}
\begin{center}
\begin{subtable}[t]{0.75\textwidth}
\centering
\begin{tabular}
{@{} M{p}{0.17} M{p}{0.19} M{p}{0.22} M{p}{0.19} @{}}
\toprule
\midrule
\textbf{YBCO}  &  \lambda      &  \tau                 & \gstar \\
\thinrule
T=0~\K\hphantom{0}%
        & 1.7\cdot10^{-7}~\mt  &  9.03\cdot10^{-11}~\s  & 2.6\cdot10^{-12}\\
T=70~\K & 2.6\cdot10^{-7}~\mt  &  2.1\cdot10^{-10}~\s   & 9.8\cdot10^{-12}\\
T=77~\K & 3.3\cdot10^{-7}~\mt  &  3.4\cdot10^{-10}~\s   & 2\cdot10^{-11} \\
T=87~\K & 8\cdot10^{-7}~\mt    &  2\cdot10^{-9}~\s      & 2.8\cdot10^{-7} \\
\midrule
\bottomrule
\end{tabular}
\smallskip
\caption{YBCO}
\label{subtab:YBCO}
\end{subtable}
\;
\begin{subtable}[t]{0.75\textwidth}
\centering
\begin{tabular}
{@{} M{p}{0.2} M{p}{0.2} M{p}{0.22} M{p}{0.21} @{}}
\toprule
\midrule
\textbf{Pb} &  \lambda            &  \tau                & \gstar \\
\thinrule
T=0~\K\hphantom{.00}%
          & 3.90\cdot10^{-8}~\mt& 1.5\cdot10^{-15}~\s & 1\cdot10^{-17}\\
T=4.20~\K & 4.3\cdot10^{-8}~\mt & 1.8\cdot10^{-15}~\s & 1.4\cdot10^{-17}\\
T=6.26~\K & 7.8\cdot10^{-8}~\mt & 6.1\cdot10^{-15}~\s & 8.2\cdot10^{-17}\\
T=7.10~\K & 2.3\cdot10^{-7}~\mt & 5.3\cdot10^{-14}~\s & 2.2\cdot10^{-15}\\
\midrule
\bottomrule
\end{tabular}
\smallskip
\caption{Pb}
\label{subtab:Pb}
\end{subtable}
%
\end{center}
\end{adjustwidth}
\label{tab:YBCOvsPb2}
\end{table}

\section{Conclusions}
It is clearly seen that $\lambda$ and $\tau$ grow with the temperature, so that one could think that the effect is maximum when the temperature is very close to the critical temperature $T_\textrm{c\,}$. However, this is true only for low-\Tct superconductors (LTSC) because in high-\Tct superconductors fluctuations are of primary importance for some Kelvin degree around $T_\textrm{c\,}$. The presence of these opposite contributions makes it possible that a temperature $\subt[T][max]\leq\Tc$ exists, at which the effect is maximum. In all cases, the time constant $\subt[t][int]$ is very small, and this makes the experimental observation rather difficult.\par
Here we suggest to use pulsed magnetic fields to destroy and restore the superconductivity within a time interval of the order
of $\subt[t][int]$. The main conclusion of this work is that the reduction of the gravitational field in a superconductor, if it exists, is a transient phenomenon and depends strongly on the parameters that characterize the superconductor.


\appendix
\addtocontents{toc}{\protect\addvspace{2.5pt}}%
\numberwithin{equation}{section}%

\section{Sign convention}\label{app:signconv}
We work in the ``mostly plus'' convention, where
\begin{equation}
\eta\=\mathrm{diag}(-1,+1,+1,+1)\;.
\end{equation}
We define the Riemann tensor as:
\begingroup
\setlength{\belowdisplayskip}{5pt plus 2pt minus 2pt}%
\begin{equation}
\begin{split}
\Ruddd&\=\dm[\lambda]\Gam[\sigma][\mu][\nu]\-\dm[\nu]\Gam[\sigma][\mu][\lambda]
        \+\Gam[\sigma][\rho][\lambda]\,\Gam[\rho][\nu][\mu]
        \-\Gam[\sigma][\rho][\nu]\,\Gam[\rho][\lambda][\mu]\=\\
      &\= 2\,\dm[{[}\lambda]\Gam[\sigma][\nu{]}][\mu]
        \+2\,\Gam[\sigma][\rho][{[}\lambda]\,\Gam[\rho][\nu{]}][\mu]\;,
\end{split}
\end{equation}
\endgroup
where
\begingroup
\setlength{\abovedisplayskip}{2pt plus 0pt minus 4pt}%
\setlength{\belowdisplayskip}{15pt plus 3pt minus 3pt}%
\begin{equation}
\begin{split}
\Gam[\lambda][\nu][\rho]&\=\invgmetr[\lambda][\mu]\,\Gamd[\mu][\nu][\rho]\;,\\
\Gamd[\mu][\nu][\rho]&\=\frac12\,\left(\dm[\rho]\gmetr[\mu][\nu]
                                      +\dm[\nu]\gmetr[\mu][\rho]
                                      -\dm[\mu]\gmetr[\nu][\rho]\right)\;.
\end{split}
\end{equation}
\endgroup
The Ricci tensor is defined as a contraction of the Riemann tensor
\begin{equation}
\Ricci\=\Ruddd[\sigma][\mu][\sigma][\nu]\;,
\end{equation}
the Ricci scalar is given by
\begin{equation}
R\=\invgmetr\Ricci\;,
\end{equation}
and the so-called Einstein tensor $\GEinst[][]$ has the form
\begin{equation}
\GEinst~\equiv~\Ricci-\dfrac12\,\gmetr\,R\;.
\end{equation}
The Einstein equations are written
\begin{equation}
\GEinst~\equiv~\Ricci-\dfrac12\,\gmetr\,R
\=8\pi\GN\;T_{\mu\nu}\;,
\end{equation}
where $T_{\mu\nu}$ is the total energy-momentum tensor. The cosmological constant contribution can be pointed out splitting $T_{\mu\nu}$ tensor in the matter and $\Lambda$ component:
\begin{equation}
T_{\mu\nu}\=T^{{}^\tts{(M)}}_{\mu\nu}+T^{{}^\ms{(\Lambda)}}_{\mu\nu}
          \=T^{{}^\tts{(M)}}_{\mu\nu}-\frac{\Lambda}{8\pi\GN}\,\gmetr\;,
\end{equation}
so that the Einstein equation can be rewritten as
\begin{equation}
\Ricci-\dfrac12\,\gmetr\,R\=8\pi\GN\,\left(T^{{}^\tts{(M)}}_{\mu\nu}+T^{{}^\ms{(\Lambda)}}_{\mu\nu}\right)\;,
\end{equation}
or, equivalently,
\begin{equation}
\Ricci-\dfrac12\,\gmetr\,R+\Lambda\,\gmetr\=8\pi\GN\;T^{{}^\tts{(M)}}_{\mu\nu}\;.
\end{equation}
%


\bigskip

\section*{\large Acknoledgments}
\vspace{-5pt}
This work was supported by the Competitiveness Program of the National Research Nuclear University MEPhI of Moscow for the contribution of prof.\ G.\ A.\ Ummarino.
We also thank Fondazione CRT \,\includegraphics[height=\fontcharht\font`\B]{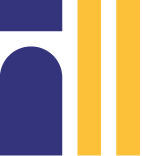}\: that partially supported this work for dott.\ A.\ Gallerati.


\newpage
\hypersetup{linkcolor=blue}
\phantomsection 
\addtocontents{toc}{\protect\addvspace{3.5pt}}
\addcontentsline{toc}{section}{References} 
\bibliographystyle{myJHEPbibstyle}
\bibliography{bibliografia} 

\end{document}